\documentclass[a4paper,12pt]{article}
\usepackage[T1,T2A]{fontenc}
\usepackage[utf8]{inputenc}
\usepackage[english]{babel}

\usepackage[a4paper]{geometry}
\usepackage{amsmath}

\usepackage{graphicx}
\usepackage[english]{babel}
\usepackage{authblk}

\title{\uppercase{ON FAST CHARGED PARTICLES SCATTERING ON A FLAT BEAM OF CHARGED  ULTRARELATIVISTIC PARTICLES IN Approximation of continuous potential}}
\author[1,2]{\fbox{N.F.~Shul'ga}$^{*,}$}

\usepackage{lipsum}

\newcommand\blfootnote[1]{%
  \begingroup
  \renewcommand\thefootnote{}\footnote{#1}%
  \addtocounter{footnote}{-1}%
  \endgroup
}

\usepackage{multicol}
\usepackage{subcaption}

\author[1,2]{V.D.~Omelchenko $^{**,}$}
\affil[1]{National Science Center ``Kharkiv Institute of Physics and Technology'', Kharkiv, Ukraine}
\affil[2]{V.N. Karazin Kharkiv National University, Kharkiv, Ukraine}
\begin{document}
\maketitle

\begin{abstract}
{The differential scattering cross section for the problem of charged ultrarelativistic particles motion near a flat beam of charged ultrarelativistic particles was obtained. The problem is considered in the eikonal approximation in the representation of the beam by a continuous potential.
\par}

\end{abstract}

\section*{{Introduction}}
\blfootnote{$^{*}$shulga@kipt.kharkov.ua, ORCID: 0000-0003-1679-6819}
\blfootnote{$^{**}$koriukina@kipt.kharkov.ua (corresponding author), ORCID: 0000-0002-5113-3028}
{The collision of relativistic beams of charged particles is of particular interest since this process often occurs in modern research at particle accelerators. The scattering process study is not only interesting in itself but also helps to examine other processes (for example, radiation) that are caused by scattering. \par}
{Using scattering theory \cite{RJG59} and quantum electrodynamics \cite{AIA65, QLP, AIA96}, we considered some problems of particle scattering on various targets from a single point of view both in the Born and in the eikonal approximations \cite{Born, Eik}. The differential scattering cross sections on targets of complex configurations were obtained in a relatively simple way, that creates an alternative to the method of simulating particle trajectory in matter. This work shows how to consider the scattering problems on moving ultrarelativistic targets without additional transformations of the coordinate system. The method demonstrated here involves the idea of a continuous potential, suggested by Lindhard \cite{JLi}. This investigation is interesting from a theoretical point of view and is also useful for the theoretical description of modern beam collision experiments held at particle accelerators, in particular at the SuperKEKB accelerator (Japan). \par}
\section{The Scattering Problem Formulation}
{Let us consider the problem of charged ultrarelativistic particles scattering  on a beam of charged ultrarelativistic particles in the eikonal approximation of quantum electrodynamics. For simplification, we will study the interaction of one incident particle with the target beam. We choose beam properties close to the properties available at modern experimental facilities, namely at the SuperKEKB accelerator (Japan) \cite{Bambade}. \par}

\begin{figure}[ht]
	\centering
			 \includegraphics[width=.8\textwidth]{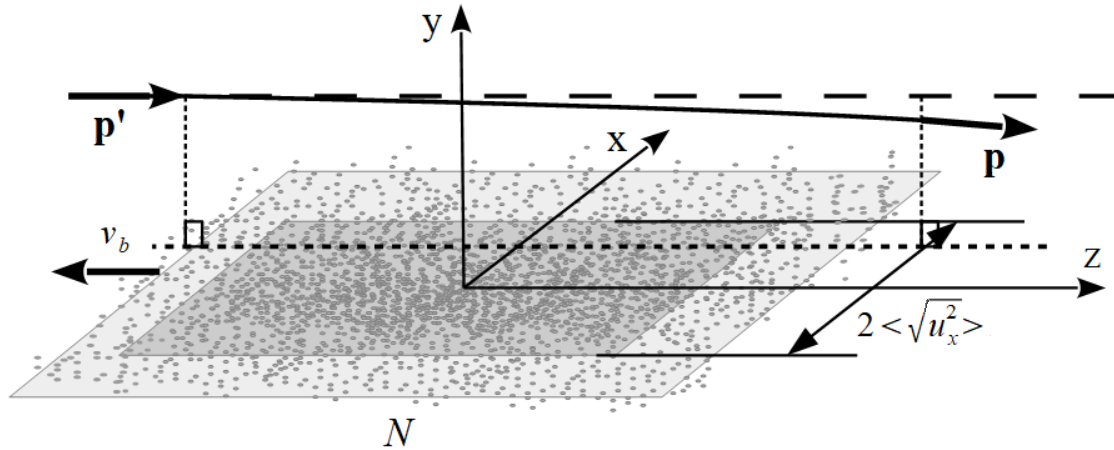}
	\caption{The motion of a fast charged particle incident parallel to the flat beam}
	\label{FIG:1}
\end{figure}

{Let the particle incident on the beam parallel to the beam velocity, along the $z$-axis. The potential of a relativistic beam is determined by the four-potential
\begin{align} \label{eq1}
    	  A_\mu^{(N)}=\left\{1,0,0,-v_b \right\} A_0^{(N)},
\end{align}
where the beam velocity has the following magnitude and direction: $\vec{v}_b=-v_b \vec{e}_z$. The total scalar potential $A_0^{(N)}$ of the target, which consists of $N$ particles of the same type (we consider electrons or positrons), is the sum of the scalar potentials of individual target particles
\begin{align} \label{eq2}
A_{0}^{(N)}=\sum_{n=1}^N A_{0, n}.
\end{align}
The scalar potential of $n$-th particle of target, in its turn, is the relativistic Coulomb potential		
\begin{align} \label{eq3}  	 
   A_{0, n}= \frac{e}{\sqrt{(z-z_n+v_bt)^2+(\vec{\rho}-\vec{\rho}_n)^2/\gamma_b^2}},
\end{align}	
where $\gamma_b$ is Lorentz factor, $z_n$ is position of the $n$-th particle along $z$-axis, $\vec{\rho}=(x,y)$ are the coordinates transverse to the beam velocity, $\vec{\rho}_n=(x_n,y_n)$ is the position of $n$-th particle of the beam relative to the center of the beam. We note that we neglect the deflection of the target due to the interaction with the incident particle, so transverse coordinates of particles of target coincide both in the coordinate system associated with the beam and in the laboratory coordinate system throughout the scattering process. \par}
{Since the beam described in \cite{Bambade} is quite flat, the particles distribution function in the target for simplification will be considered as follows: along the $x$-axis, the beam particles are distributed according to the Gaussian law with a random deviation $x_n$ and variance $\langle u^2_x \rangle$; along the $y$-axis, the beam has a zero thickness, which is equivalent to a zero variance $\langle u^2_y \rangle$:
\begin{align} \label{eq4}
	 \vec{\rho}_n: \begin{cases} f(x_n)=\frac{1}{\sqrt{2 \pi \langle u^2_x \rangle}} \: exp \left[\frac{-x^2_n}{2 \langle u^2_x \rangle} \right], \\
		y_n=0, \: \langle u^2_y \rangle =0. \end{cases}	
\end{align}

	\section{The Wave Function of a Scattered Particle}
{From the Klein-Gordon and Dirac equations, we find in the eikonal approximation the wave function of a particle being scattered on an ultrarelativistic beam. The Klein-Gordon equation for a particle of mass $m$, four-momentum $p_\mu$, moving in a four-potential $A_\mu^{(N)}$ is
\begin{align} \label{eq5}
	\left[ \left( p_\mu -eA_\mu^{(N)} \right)^2 - m^2 \right] \Phi=0.
\end{align}     	
The wave function $\Phi$ in the eikonal approximation is \cite{AIA96}
\begin{align} \label{eq6}
\Phi = f \: e^{iS}.	
\end{align} 
The amplitude $f$ and the phase $S$ of this wave function can be expanded into a series of the reciprocal powers of the momentum $f_k \sim p^{-k}, \chi_k^{(N)} \sim p^{-k}$ respectively:
\begin{align} \label{eq7}
f = f_0 +f_1(\vec{r},t)+... \ ,	\\
S = \vec{p} \vec{r} -Et + \chi_0^{(N)}(\vec{r},t)+ \chi_1^{(N)}(\vec{r},t)+... \ .
\end{align}
Let us determine $S$ through the relativistic Hamilton-Jacobi equation \cite{LLt2}
\begin{align} \label{eq9}
\left( \partial_t S +eA_0^{(N)} \right)^2-\left(\nabla S -e\vec{A}^{(N)} \right)^2 - m^2 =0.	
\end{align}
It leads to the equation
\begin{align} \label{eq10}
 \partial_z \chi_0^{(N)} + \partial_t \chi_0^{(N)} + 2 eA_0^{(N)} =0,	
\end{align}
from which the  $\chi_0$-function for scattering on a target of $N$ ultrarelativistic particles is
\begin{align} \label{eq11}
\chi_0^{(N)}= - \int_{-\infty}^{\xi} d \xi eA_0^{(N)}(\vec{\rho}, \xi),	
\end{align}
where $\xi=z+v_b t$, assuming that $v_b$ is approximatly equal to speed of light. So, from the Dirac equation, we obtaine the solution, which satisfies equations \eqref{eq5} and \eqref{eq9}:
\begin{align} \label{eq12}
f_0=u,	
\end{align}
where $u$ is a bispinor. \par}
{Hence, the wave function for the given case in the zero eikonal approximation is
\begin{align} \label{eq12}
\Phi=u e^{i \left[\vec{p} \vec{r} -Et + \chi_0^{(N)}(\vec{\rho},z+v_b t) \right]}.
\end{align}
\par}

\section{The Scattering Cross Section in the Eikonal Approximation }
{Knowing the general form of the wave function for the scattering of a charged ultrarelativistic particle on a beam of charged ultrarelativistic particles, we will find the general form of the differential scattering cross section for this problem. In accordance with quantum electrodynamics, the differential scattering cross section for a particle, which from the initial state with energy $\varepsilon '$ and momentum $\vec{p} \: '$ transits to the final state with energy $\varepsilon$ and momentum $\vec{p} $, is defined as \cite{AIA65, QLP}
\begin{align} \label{eq13}
d \sigma = \frac{1}{j} \frac{w_{fi}}{\tau} \frac{V \: d^3 p }{(2 \pi)^3},	
\end{align}
where $w_{fi}$ is the probability of a particle transition from the initial state to the final state, $\tau$ is the observation time of the scattering process, $V$ is the normalizing volume, $j$ is the flow density of interacting particles, which is expressed as
\begin{align} \label{eq14}
j=\frac{2v}{V},	
\end{align}
where $v$ is the interacting particles velocity, assuming that the particles move with the same velocity magnitudes. The probability of the particle transition from the initial state to the final state is given as follows
\begin{align} \label{eq15}
w_{fi}=\frac{\left|M_{fi} \right|^2}{(2 \varepsilon V)(2 \varepsilon ' V)}	,
\end{align}
\begin{align}\label{eq16}
M_{fi}= \int  dt d^3r \:    \bar{\Phi} (\vec{r},t) (\gamma_0+\gamma_3 v) eA_0^{(N)}(\vec{\rho}, z+vt) u' e^{i (-\varepsilon ' t + \vec{p}' \vec{r})},	
\end{align}
where $\bar{u}'$ is the bispinor of the initial state, $\gamma_0$ and $\gamma_3$ are the Dirac matrices, $\Phi$ is the wave function of the final state in the zero eikonal approximation \eqref{eq12}. \par}
{Considering that for ultrarelativistic particles $\varepsilon \rightarrow pc$, we obtain 
\begin{align}\label{eq18}
M_{fi} \approx 2 \bar{u} \gamma_0 u'   \int dt d\xi d^2\rho \: e^{i (\varepsilon - \varepsilon' + q_z v  )t - i \vec{q}_\perp \vec{\rho}-i q_z \xi} eA_0^{(N)}(\vec{\rho}, \xi) \: e^{-i \chi_0 (\vec{\rho},\xi)}	
\end{align}
where  $\vec{q}=\vec{p}-\vec{p} \: '$ is momentum transfer: $\vec{q}=(\vec{q}_{\perp},q_z)$, where $q_z \ || \ \vec{e}_z$, $\vec{q}_{\perp} \perp \vec{e}_z$.\par}

\section{Discussion of the Violation of the Energy Conservation Law }
{Integrating \eqref{eq18} over $t$ leads to the emergence of the Dirac delta function $\delta(\varepsilon ' - \varepsilon - q_z v)$, which in the stationary target case is $\delta(\varepsilon ' - \varepsilon)$ and corresponds to the energy conservation law. The "new" energy conservation law has no physical sense. Let us consider this expression in detail. \par}
{ According to the energy-momentum conservation laws, the momentum transferred along the motion direction is expressed with some accuracy as 
\begin{align} \label{eq19}
q_z  \approx \frac{\varepsilon^2 - \varepsilon^{'2} +\vec{q}^2_{\perp}}{2p},	
\end{align}
where $\vec{q}_{\perp}$ is momentum transfer in perpendicular to the movement direction. Then
\begin{align} \label{eq20}
\varepsilon ' - \varepsilon - q_z v \approx 2 (\varepsilon - \varepsilon ').	
\end{align}
As a result, after substituting \eqref{eq20}, the formula \eqref{eq18} becomes 
\begin{align} \label{eq21}
M_{fi} \approx 2 \bar{u} \gamma_0 u'   \int  dt d\xi d^2 \rho \: e^{2i (\varepsilon - \varepsilon')t - i \vec{q}_\perp \vec{\rho} -iq_z \xi} eA_0^{(N)}(\vec{\rho}, \xi) \:e^{-i \chi_0 (\vec{\rho},z) },	
\end{align}
which corresponds to $M_{fi}$ in the zero eikonal approximation. In this from, the energy conservation law is fulfilled.
\par}
{Beyond the zero eikonal approximation, another way to "restore" the energy conservation law can be demonstrated. In the first eikonal approximation
\begin{align} \label{eq22}
M_{fi} \approx 2   \int  dt d^2 \rho d\xi \ (\bar{u}+\bar{f}_1) \gamma_0 u' e^{i (\varepsilon - \varepsilon' + q_z v  )t - i \vec{q}_\perp \vec{\rho} -iq_z \xi} eA_0^{(N)}(\vec{\rho}, \xi)  e^{-i \chi_0 (\vec{\rho},\xi) - i \chi_1 (\vec{\rho},\xi)},	
\end{align}
where $\chi_1 (\vec{\rho},\xi)$ from the Hamilton-Jacobi equation \eqref{eq9} is
\begin{align}\label{eq23}
\chi_1 = - \frac{1}{4p} \int^{\xi}_{-\infty} d\xi \ \left( \partial_{\vec{\rho}} \chi_0 \right)^2.
\end{align}
At a certain choice of the coordinate-time system: $z=vt$, then the integration with given limits can be replaced by integration over the region where the particle interacts with the beam, that is, from $\xi=z+vt=0$ to $\xi=z+vt=2L_z$, where $ L_z$ is the longitudinal size of the beam. Then, using the method of integration by parts, we get
\begin{align}\label{eq24}
\chi_1 =- \frac{\xi \left( \partial_{\vec{\rho}} \chi_0 \right)^2 \Big|^{\xi=2L_z}_{\xi=0}}{4p} + \frac{1}{4p} \int^{\xi}_{-\infty} d\left( \partial_{\vec{\rho}} \chi_0 \right)^2 \ \xi.
\end{align}
Considering that in accordance with the quasi-classical approximation $|\partial_{\vec{\rho}} \chi_0| = {q}_{\perp}$, the first term in \eqref{eq24} becomes $\frac{{ q}_{\perp}^2 L_z}{2p}$ and is equivalent to the term $q_z v t$. It follows that the term $q_z v t$ vanishes due to the redefinition of the $\chi_1 (\vec{\rho},\xi)$-function  and the energy conservation law  is fulfilled. \par}
{Integrating \eqref{eq21} over $t$ and $\xi$, we obtaine the formula for the differential scattering cross section, which coincides with the corresponding formula for scattering on a stationary target
\begin{align}\label{eq25}
\frac{d^2\sigma}{dq_x dq_y}=|a(\vec{q}_\perp)|^2,	
\end{align}
where
\begin{align}\label{eq26}
a(\vec{q}_\perp)=\int _{-\infty}^{\infty} d^2 \rho e^{\frac{i}{\hbar} \vec{q} \vec{\rho} } \left\{ 1- exp \left[\frac{i}{\hbar}  {\chi}_0^{(N)}\left( \vec{\rho} \right)  \right] \right\}.
\end{align}
where $\chi_0^{(N)}$ is redefined as
\begin{align} \label{eq26_1}
\chi_0^{(N)}= -\int_{-\infty}^{\infty} d \xi eA_0^{(N)}(\vec{\rho}, \xi).	
\end{align}
Using the method described in \cite{RJG59, Eik} with accuracy up to linear in  $\bar{\chi}^{(1)}_0$ terms ($\bar{\chi}^{(1)}_0$ is ${\chi}^{(1)}_0$ averaged over the particles positions in the target and reduced to one particle of the target so that $\bar{\chi}^{(N)}_0=N\bar{\chi}^{(1)}_0$) we obtain the formula for the differential scattering cross section
\begin{align}\label{eq27}
\langle \frac{d^{2} \sigma }{dq_{\perp}^{2}} \rangle \approx \frac{1}{4 \pi ^{2}} \left|  \int d^{2} \rho e^{\frac{i }{\hbar} \left[  \vec{q}_{\perp}  \vec{\rho} +N \bar{\chi}^{(1)}_0 \left(   \vec{\rho} \right) \right] } \right|^2.	
\end{align}
The formula \eqref{eq27} depends only on variables which remain unchanged during Lorentz transformations so the given cross section is invariant. Thus, the coincidence of differential cross sections for ultrarelativistic and stationary targets is justified. We note that neglecting quadratic in $\bar{\chi}^{(1)}_0$ terms simplifies the problem consideration and corresponds to the continuous potential approximation.
 \par}

\section{The Differential Cross Section for a Particle Scattering on a Flat Beam}
{Let us obtain the necessary functions for the scattering cross section calculation, namely $\bar{\chi}_0^{(1)}$ and its derivatives. Averaging ${\chi}_0^{(1)}$ using the particle distribution function in the target \eqref{eq4} gives
\begin{align}\label{eq28}
\bar{\chi}_0^{(1)} = 2 \alpha Re \left\{ \frac{(|y|+ix)^2}{2\langle{u_x^2} \rangle} \:_2F_2 \left[1,1;\frac{3}{2},2; \frac{(|y|+ix)^2}{2\langle{u_x^2} \rangle} \right] -\frac{\pi}{2}Erfi \left[ \frac{|y|+ix}{\sqrt{2\langle{u_x^2} \rangle}} \right] \right\}  +  const,	
\end{align}
where $\alpha$ is the fine structure constant, $\:_2F_2(a_1,a_2;b_1,b_2;x)$ is generalized hypergeometric function, $Erfi(x)$ is imaginary error function, $const$ is determined by the cutoff which we need to make, since the long-range potential of the beam leads to the divergence of the $\bar{\chi}_0^{(1)}$-function	.
		
\begin{figure}
	\centering
			 \includegraphics[width=.6\textwidth]{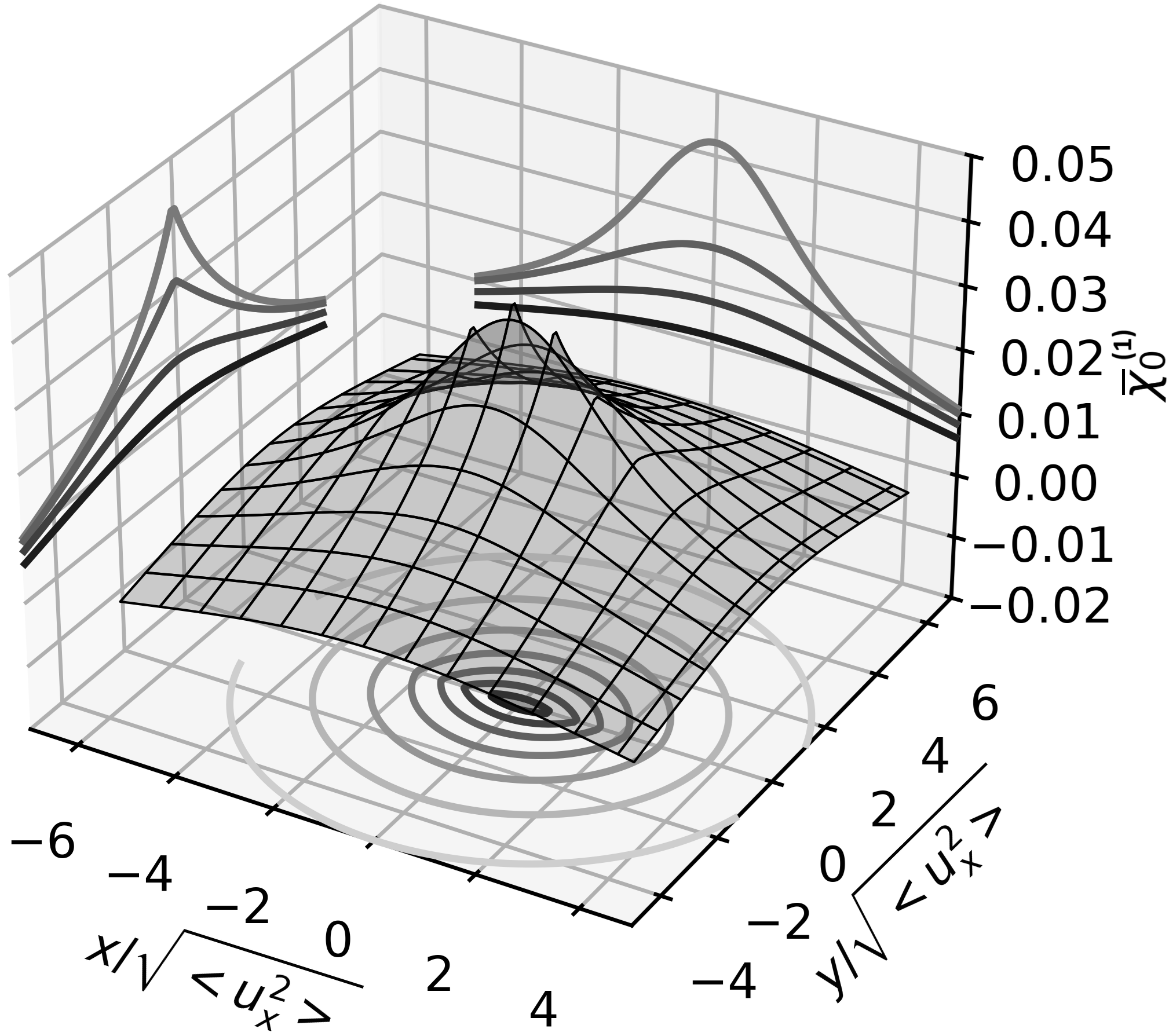}
	\caption{The $\bar{\chi}_0^{(1)}$-function for the flat beam}
	\label{FIG:2}
\end{figure}
\begin{figure}[ht]
      \centering
	   \begin{subfigure}{0.49\linewidth}
		\includegraphics[width=\linewidth]{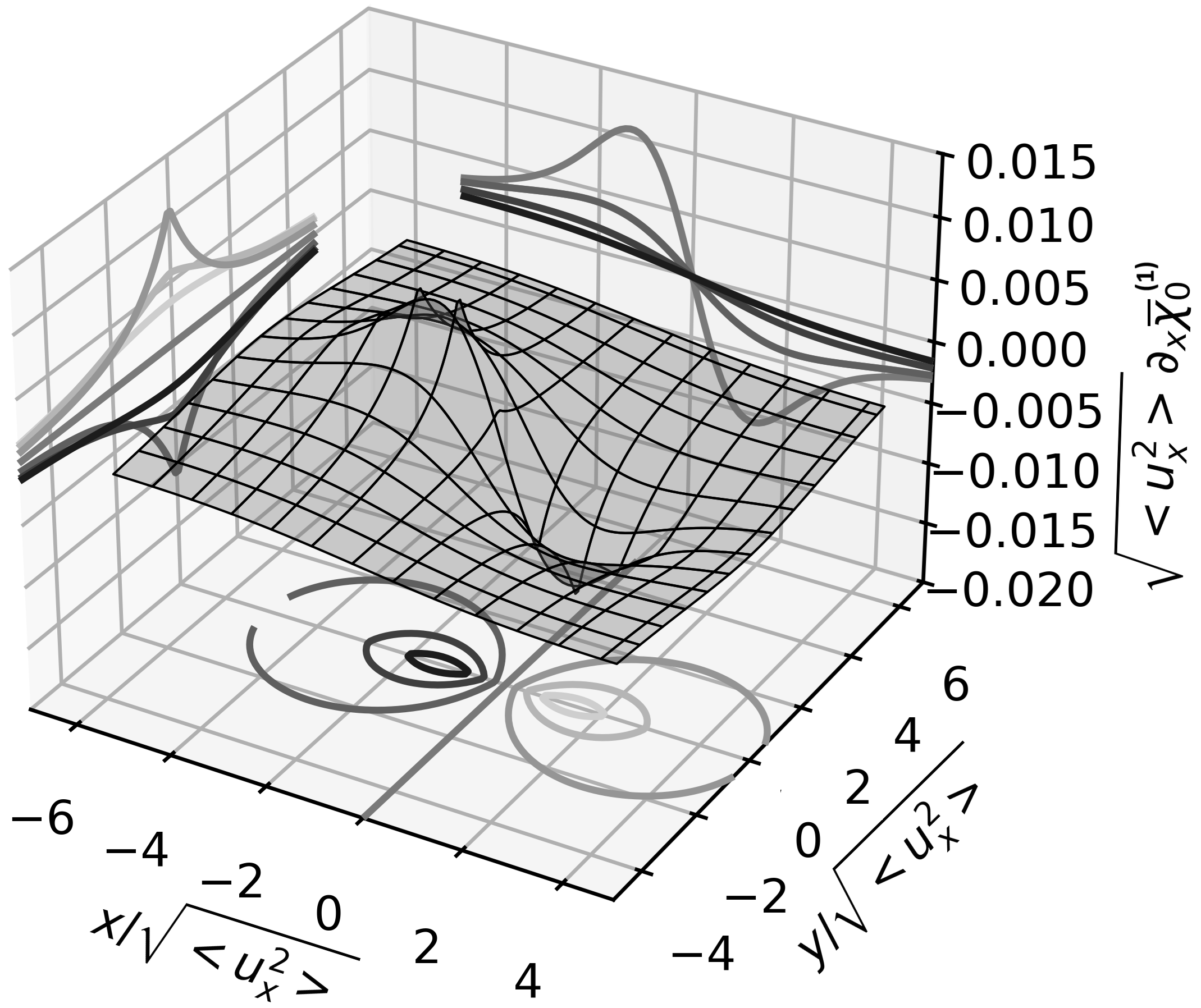}
		\caption{$\partial_x \bar{\chi}_0^{(1)}$}
		\label{fig:subfig_1}
	   \end{subfigure}
	   \begin{subfigure}{0.49\linewidth}
		\includegraphics[width=\linewidth]{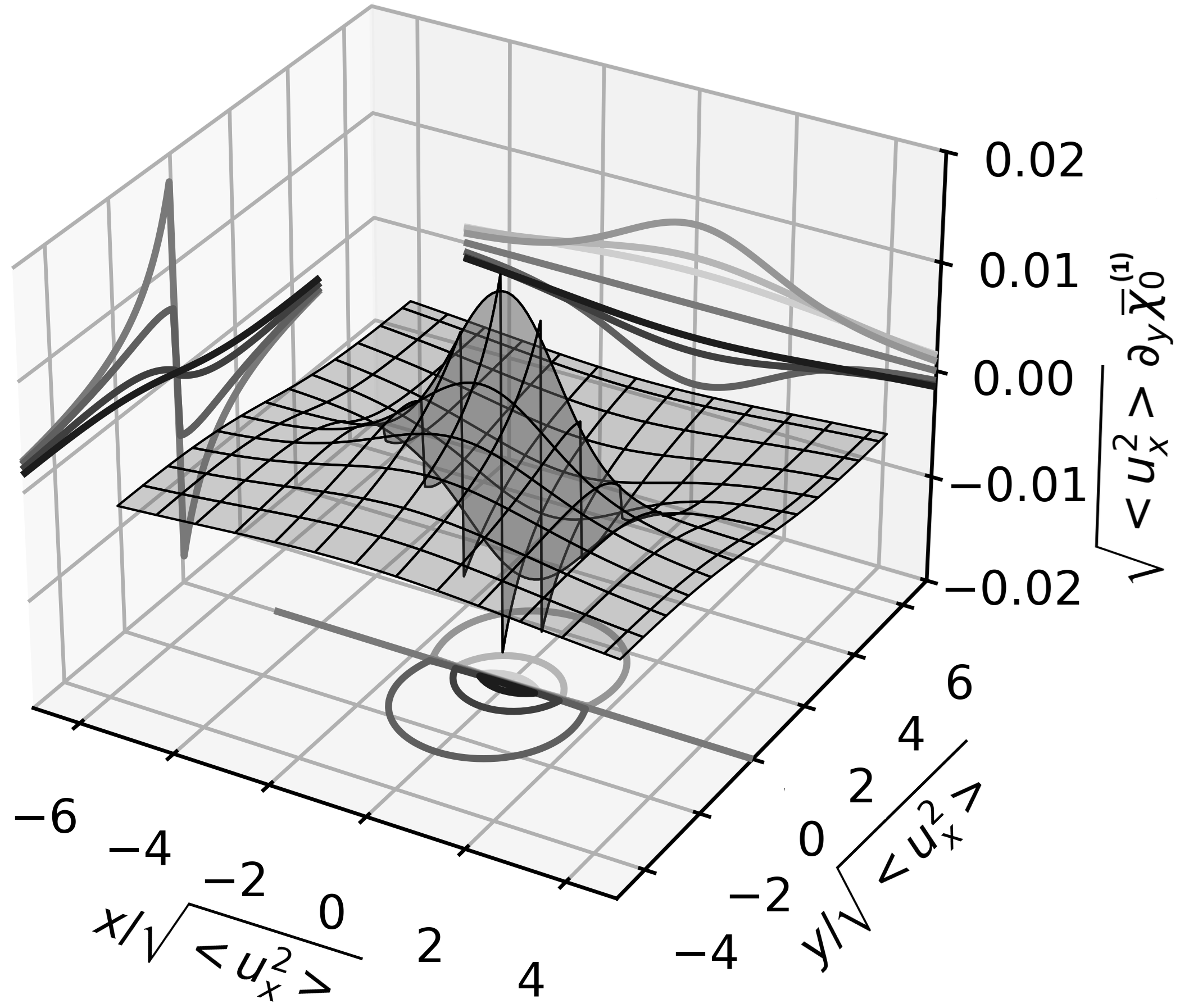}
		\caption{$\partial_y \bar{\chi}_0^{(1)}$}
		\label{fig:subfig_2}
	    \end{subfigure}
	\caption{Derivatives of the $\bar{\chi}_0^{(1)}$-function for the flat beam }
	\label{FIG:3}
\end{figure}
\begin{figure}[ht]
      \centering
	   \begin{subfigure}{0.49\linewidth}
		\includegraphics[width=\linewidth]{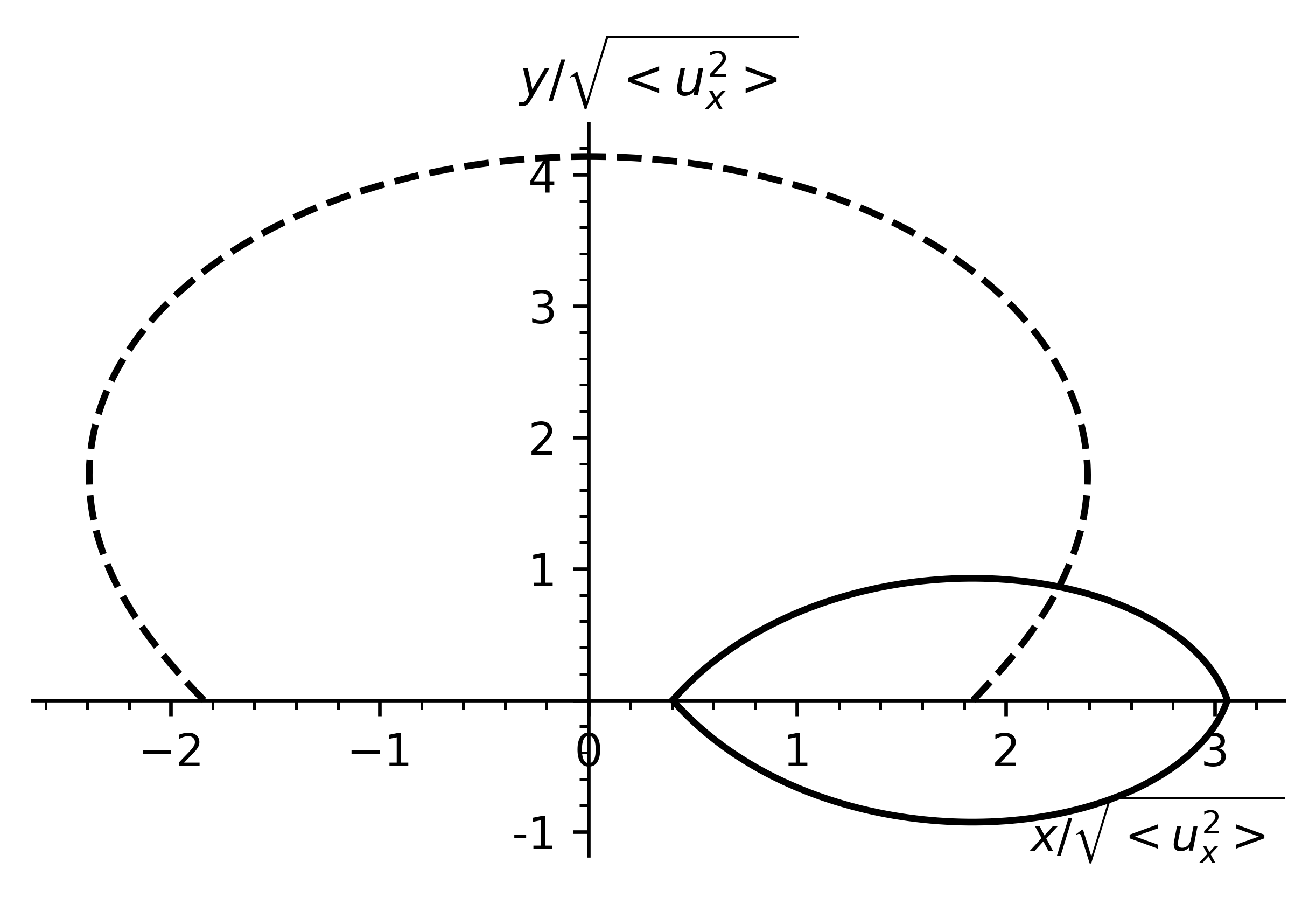}
		\caption{$q_x=0.5 \ q_{x \ max}, \ q_y=0.18 \ q_{y \ max}$}
		\label{fig:subfig0_1}
	   \end{subfigure}
	   \begin{subfigure}{0.49\linewidth}
		\includegraphics[width=\linewidth]{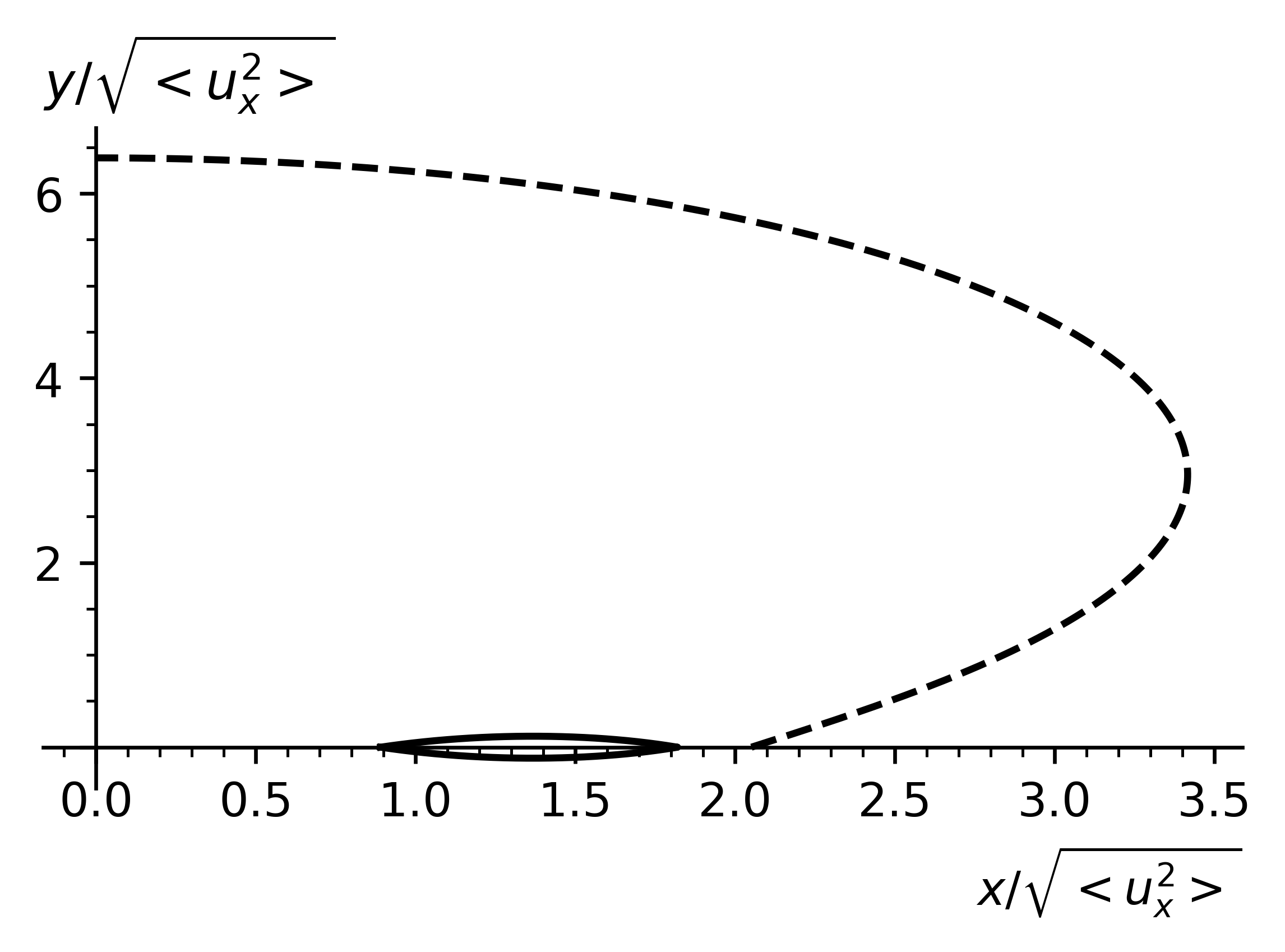}
		\caption{$q_x=0.9 \ q_{x \ max}, \ q_y=0.12 \ q_{y \ max}$}
		\label{fig:subfig0_2}
	    \end{subfigure}
	\caption{Graphic solution of \eqref{eq32} for specific values of momentum transfer $\left( q_x, q_y \right)$: solid line corresponds to $\partial_x \bar{\chi}_0^{(1)} =const$, dashed line -- to $\partial_y \bar{\chi}_0^{(1)} =const$ }
	\label{fig:0}
\end{figure}
\begin{figure}[ht]
      \centering
	   \begin{subfigure}{0.49\linewidth}
		\includegraphics[width=\linewidth]{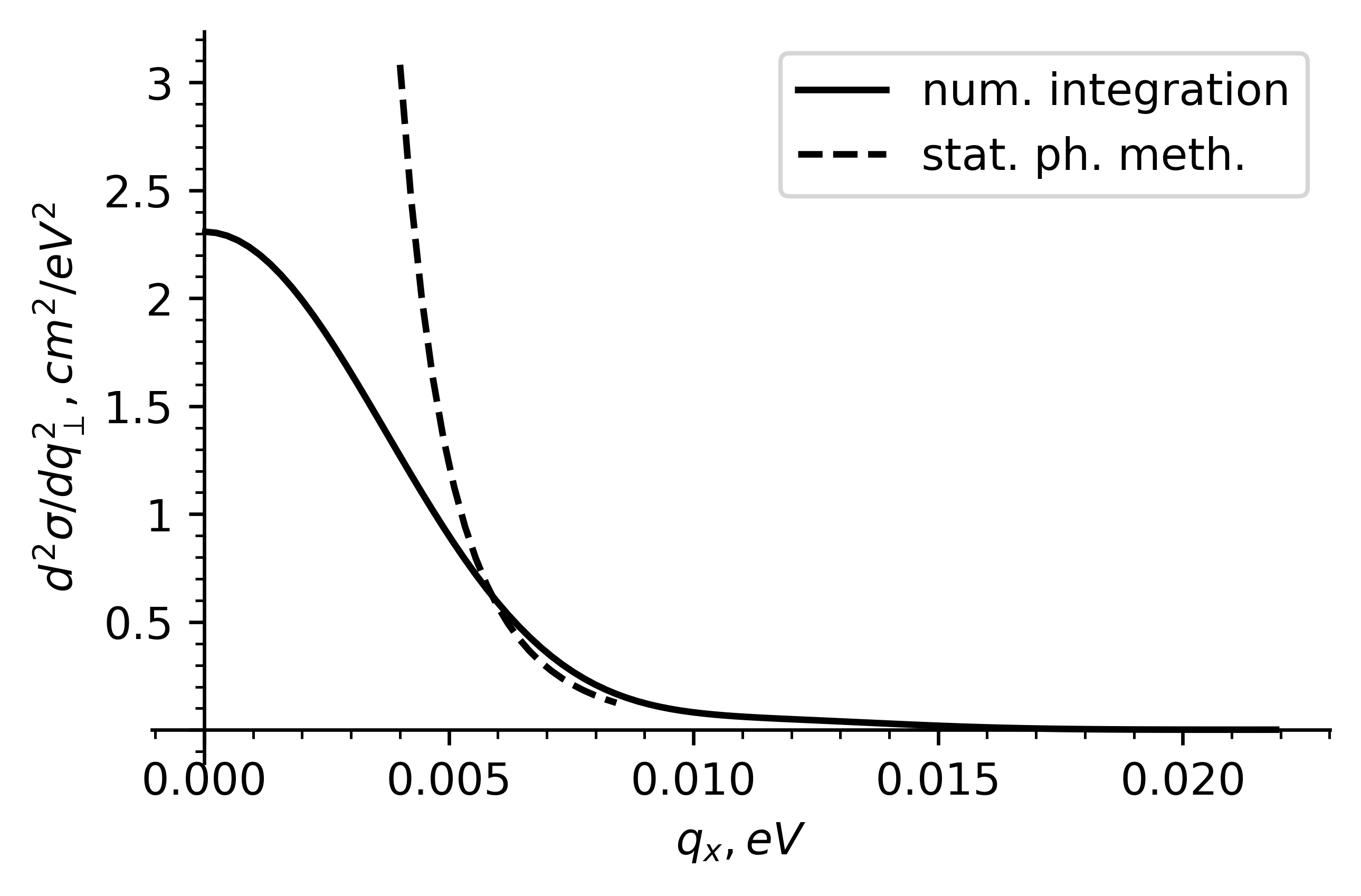}
		\caption{$q_y=3.6 \ 10^{-5} \ eV$}
		\label{fig:subfig1_1}
	   \end{subfigure}
	     \begin{subfigure}{0.49\linewidth}
		 \includegraphics[width=\linewidth]{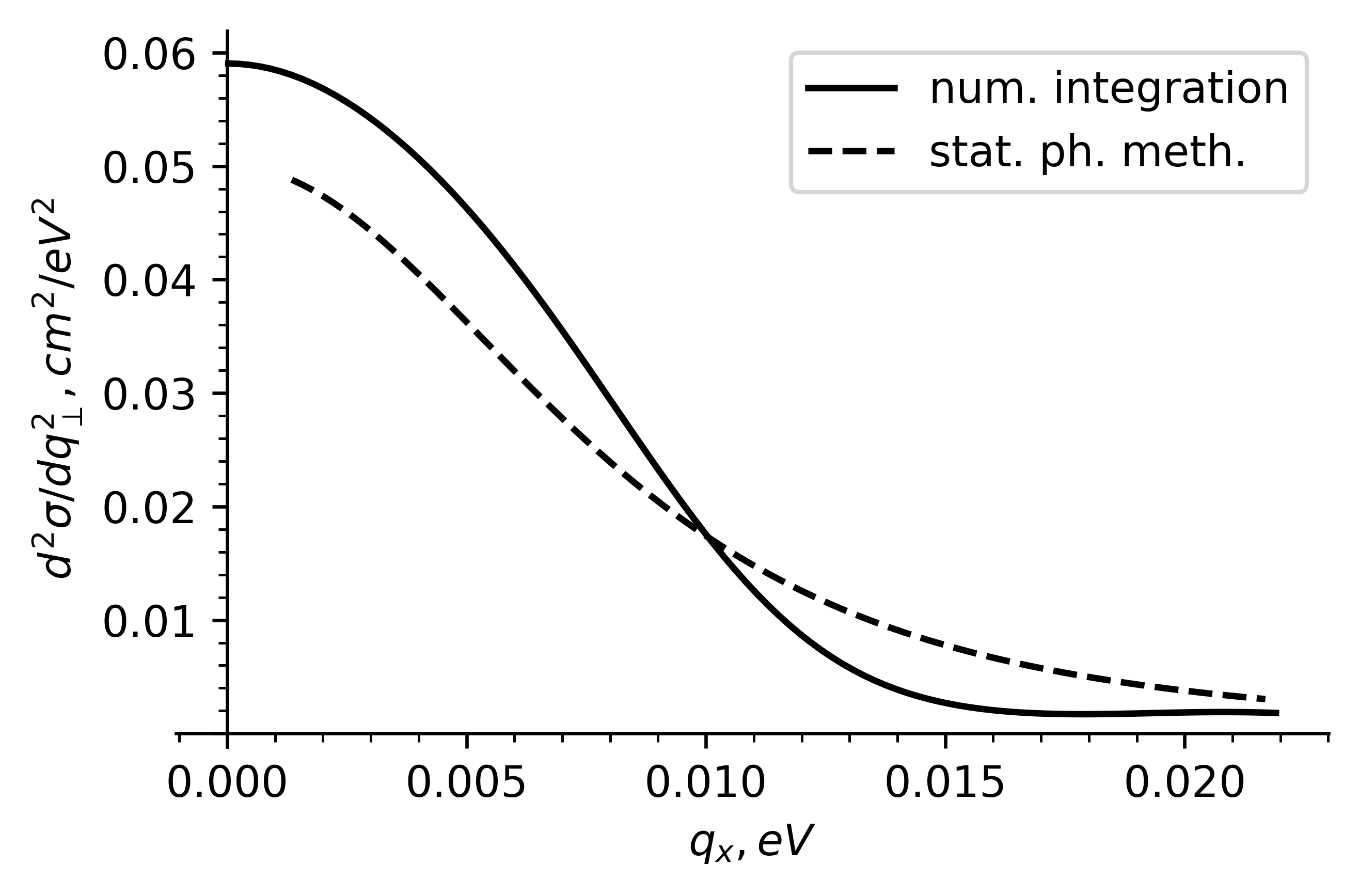}
		 \caption{$q_y=0.012 \ eV$}
		 \label{fig:subfig1_2}
	      \end{subfigure}
	\vfill
\begin{subfigure}{0.49\linewidth}
		\includegraphics[width=\linewidth]{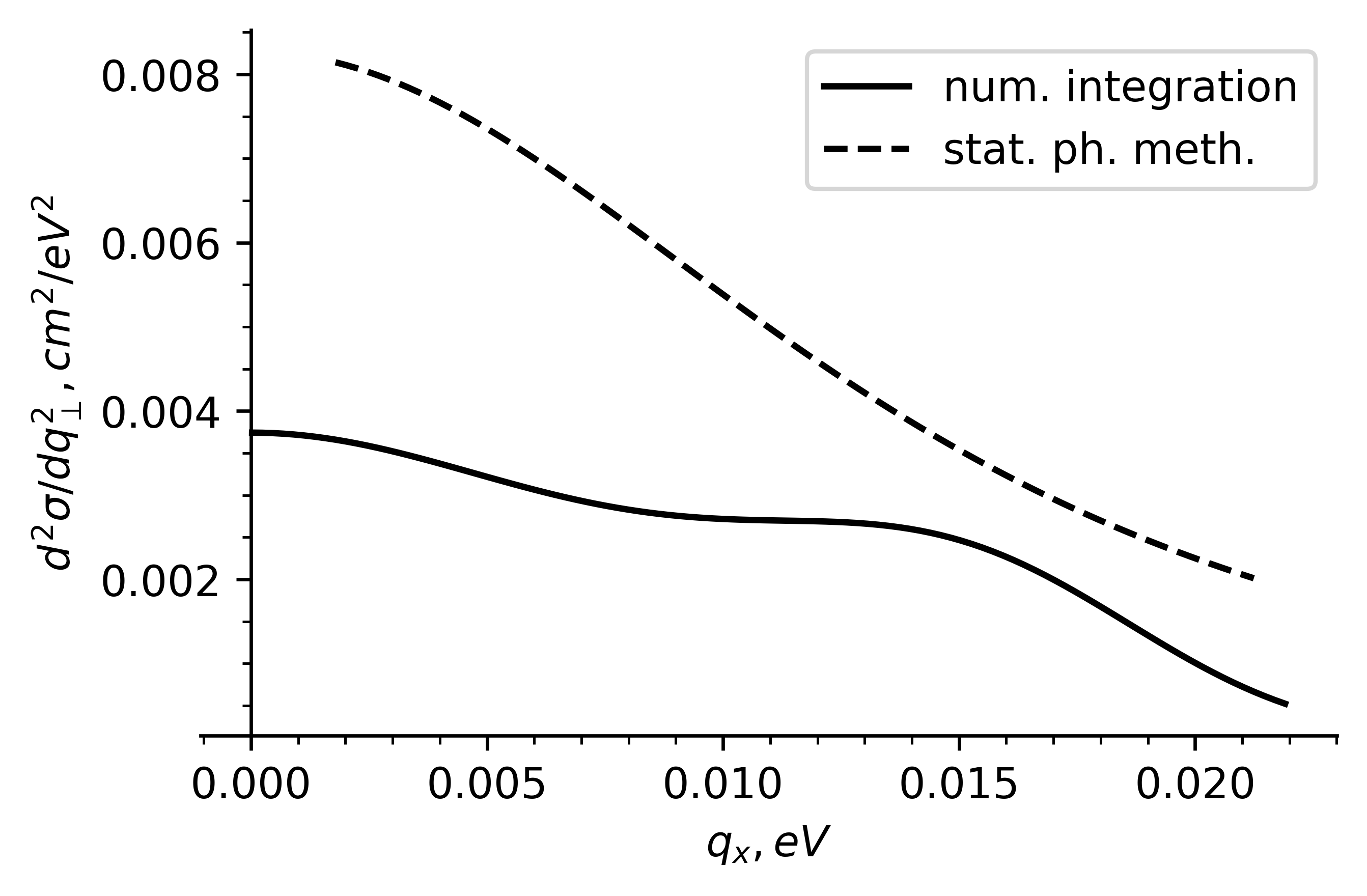}
		\caption{$q_y=0.02 \ eV$}
		\label{fig:subfig1_3}
	    \end{subfigure}
	       \begin{subfigure}{0.49\linewidth}
		  \includegraphics[width=\linewidth]{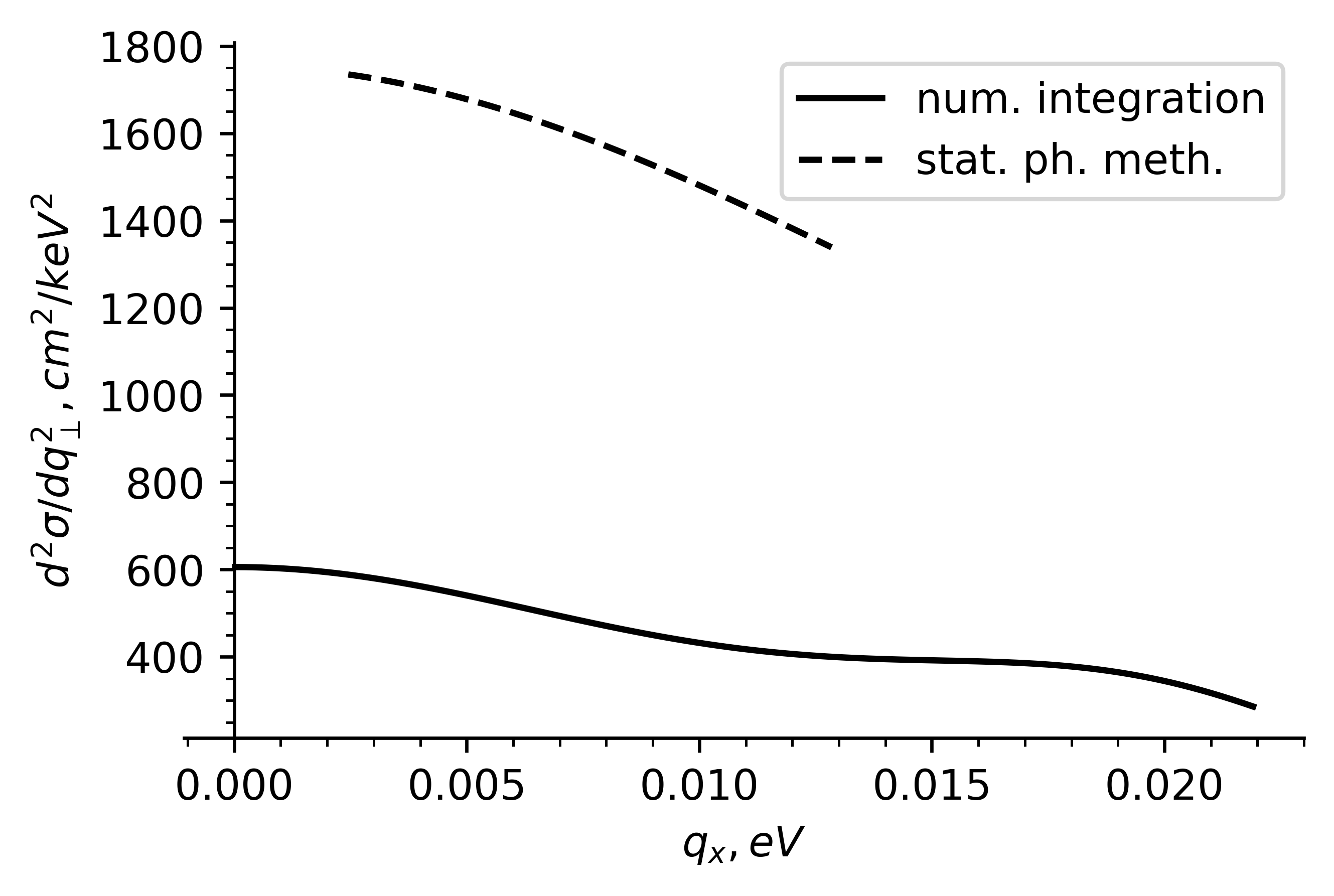}
		  \caption{$q_y=0.032 \ eV$}
		  \label{fig:subfig1_4}
	       \end{subfigure}
	\caption{The differential cross section for scattering on the flat beam of $N=100$ particles and various $q_y$ values: solid line corresponds to numerical integration in \eqref{eq27}, dashed line -- to stationary phase method \eqref{eq31}}
	\label{fig:1}
\end{figure}
\begin{figure}[!ht]
      \centering
	   \begin{subfigure}{0.49\linewidth}
		\includegraphics[width=\linewidth]{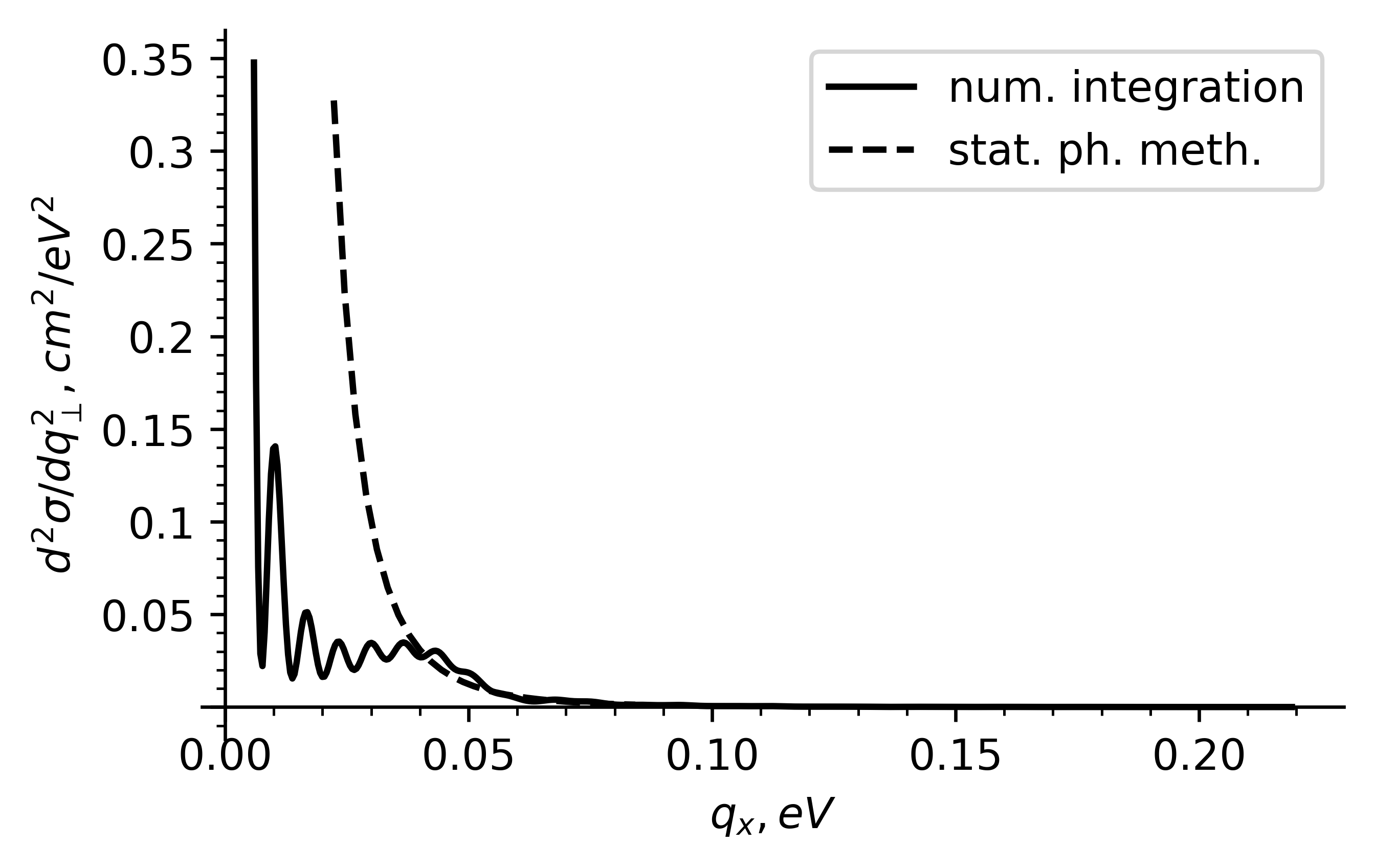}
		\caption{$q_y=3.6 \ 10^{-4} \ eV$}
		\label{fig:subfig2_1}
	   \end{subfigure}
	   \begin{subfigure}{0.49\linewidth}
		\includegraphics[width=\linewidth]{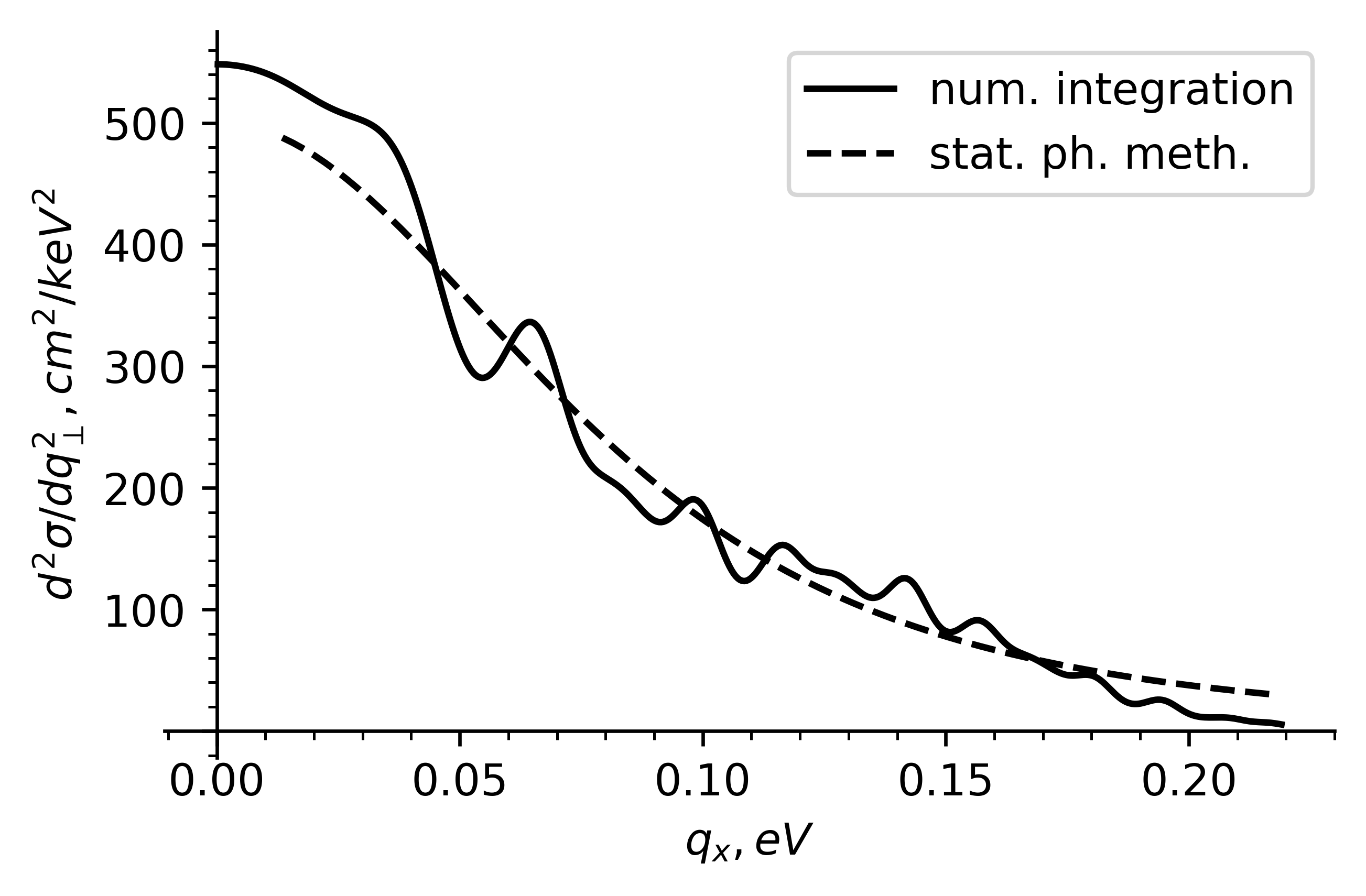}
		\caption{$q_y=0.12 \ eV$}
		\label{fig:subfig2_2}
	    \end{subfigure}
	\vfill
	     \begin{subfigure}{0.49\linewidth}
		 \includegraphics[width=\linewidth]{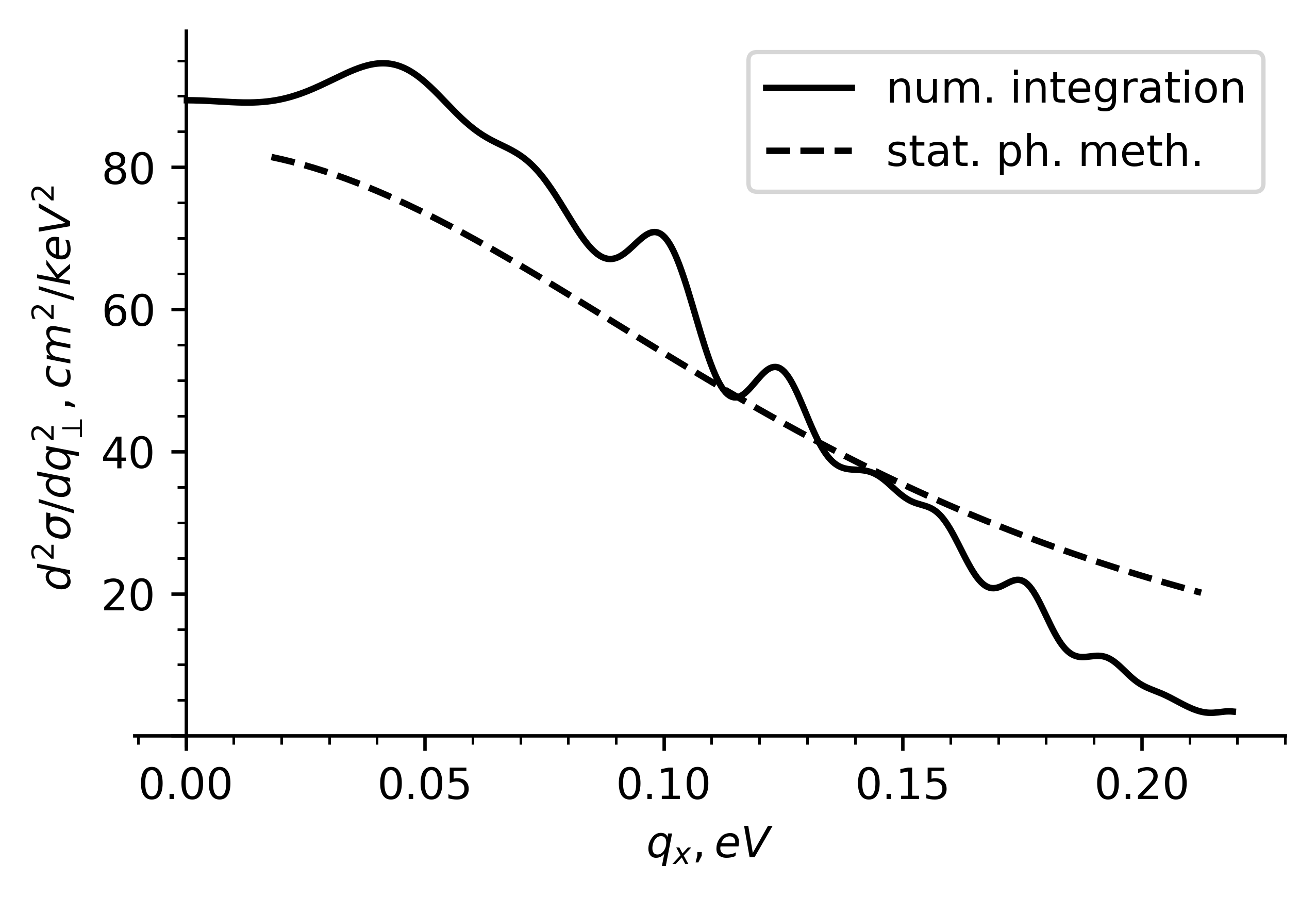}
		 \caption{$q_y=0.2 \ eV$}
		 \label{fig:subfig2_3}
	      \end{subfigure}
	       \begin{subfigure}{0.49\linewidth}
		  \includegraphics[width=\linewidth]{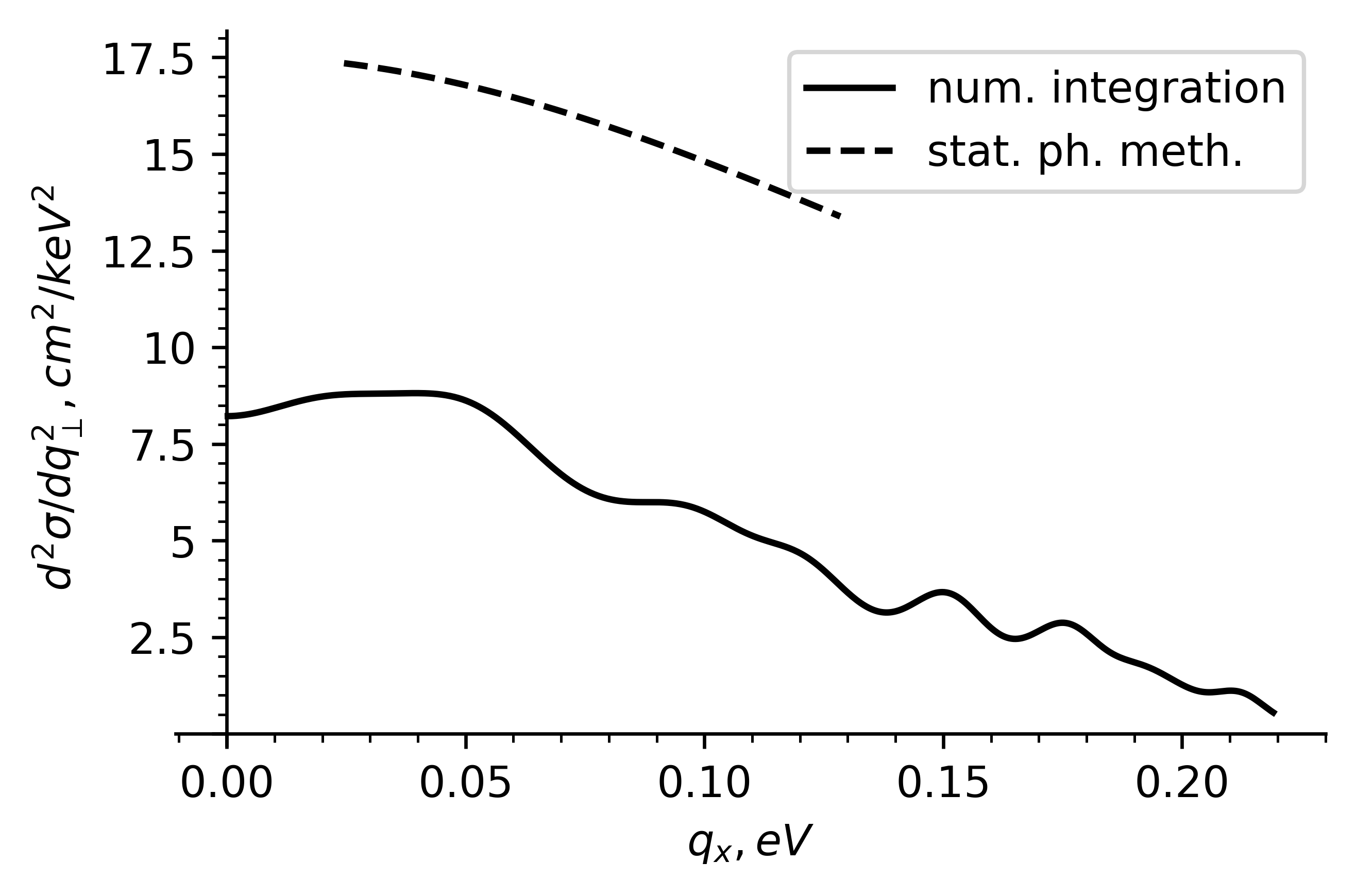}
		  \caption{$q_y=0.32 \ eV$}
		  \label{fig:subfig2_4}
	       \end{subfigure}
	\caption{The differential cross section for scattering on the flat beam of $N=1\ 000$ particles and various $q_y$ values: solid line corresponds to numerical integration in \eqref{eq27}, dashed line -- to stationary phase method \eqref{eq31}}
	\label{fig:2}
\end{figure}
\begin{figure}[!ht]
      \centering
	   \begin{subfigure}{0.49\linewidth}
		\includegraphics[width=\linewidth]{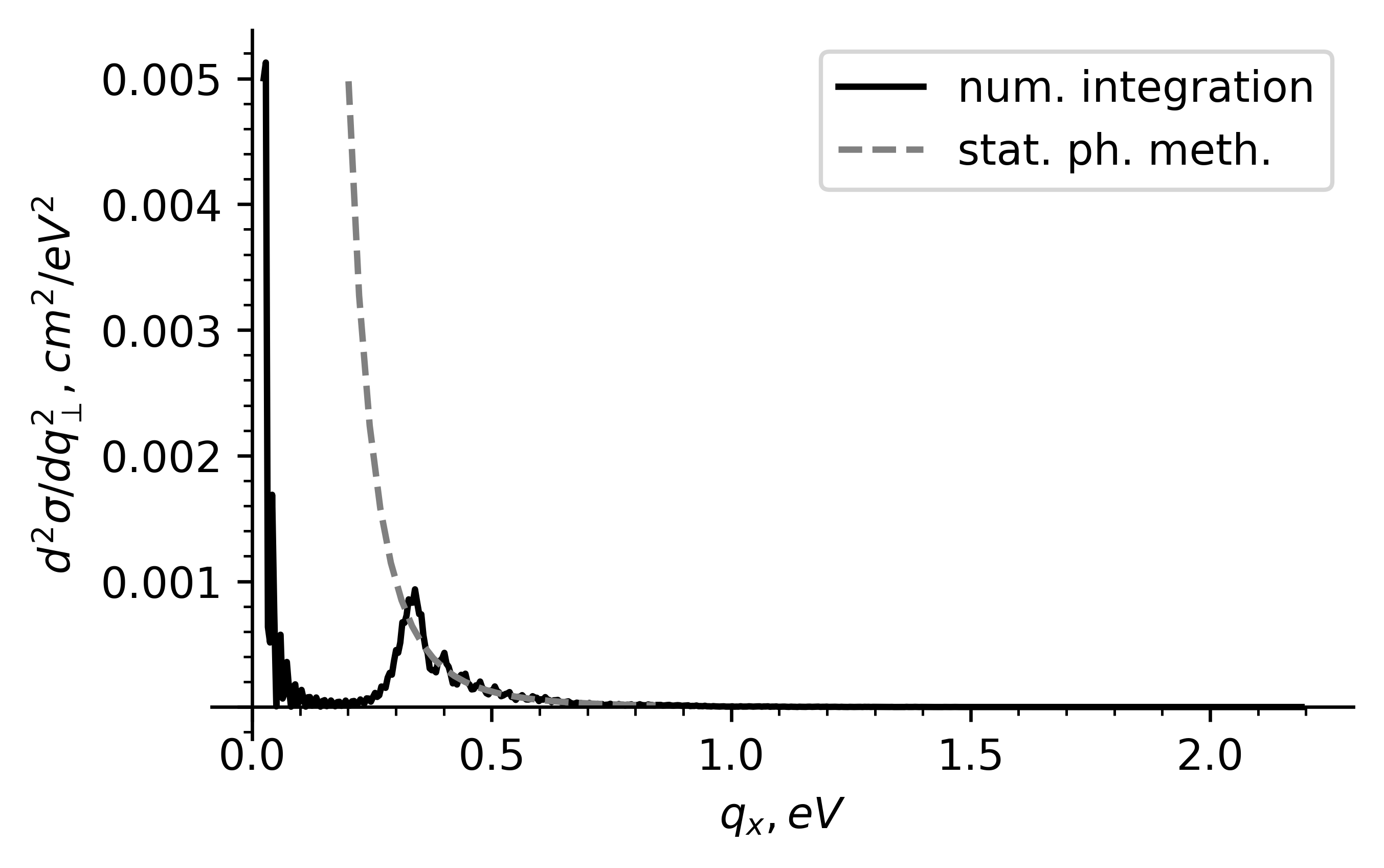}
		\caption{$q_y=3.6 \ 10^{-3} \ eV$}
		\label{fig:subfig3_1}
	   \end{subfigure}
	   \begin{subfigure}{0.49\linewidth}
		\includegraphics[width=\linewidth]{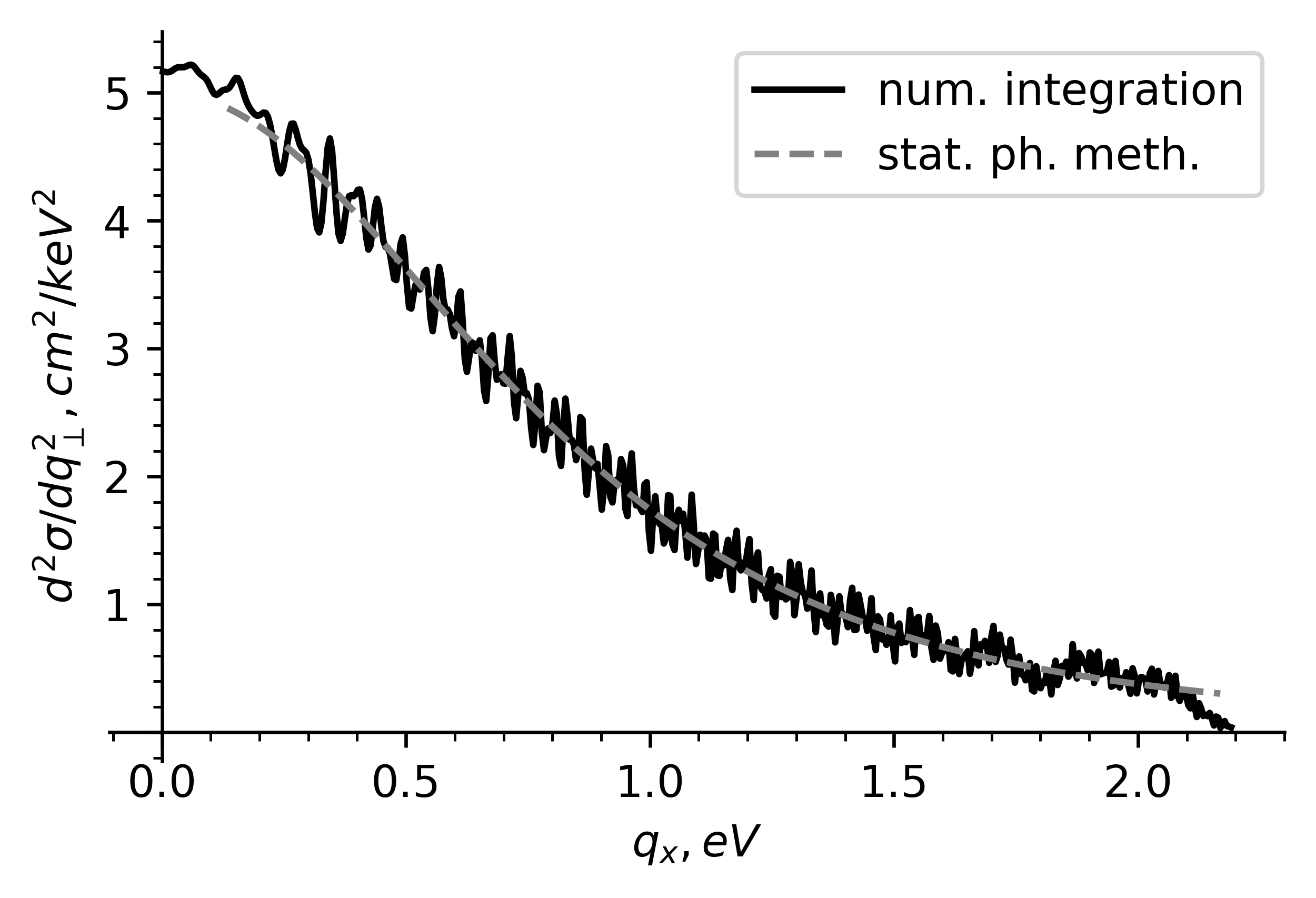}
		\caption{$q_y=1.2 \ eV$}
		\label{fig:subfig3_2}
	    \end{subfigure}
	\vfill
	     \begin{subfigure}{0.49\linewidth}
		 \includegraphics[width=\linewidth]{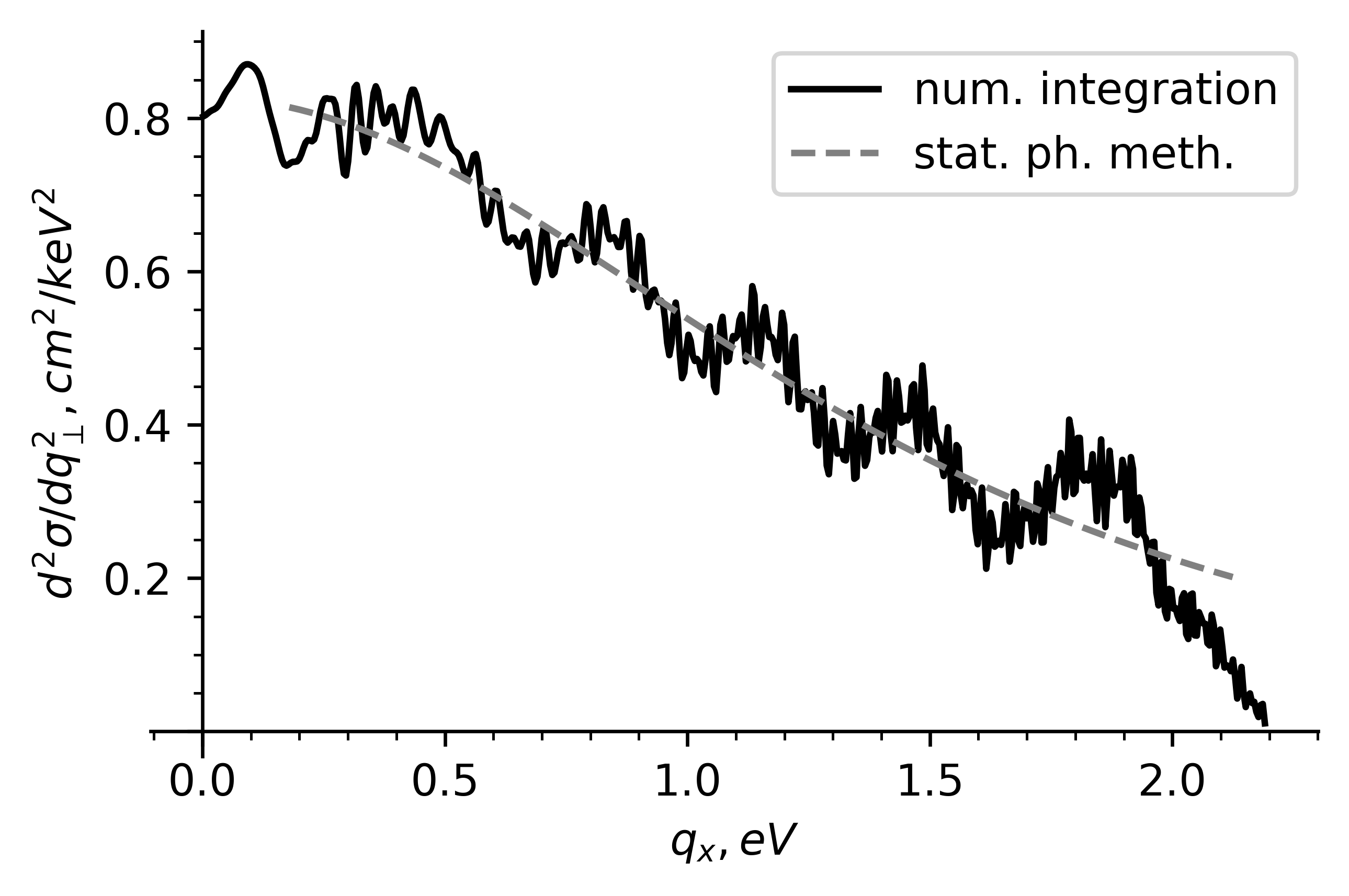}
		 \caption{$q_y=2 \ eV$}
		 \label{fig:subfig3_3}
	      \end{subfigure}
	       \begin{subfigure}{0.49\linewidth}
		  \includegraphics[width=\linewidth]{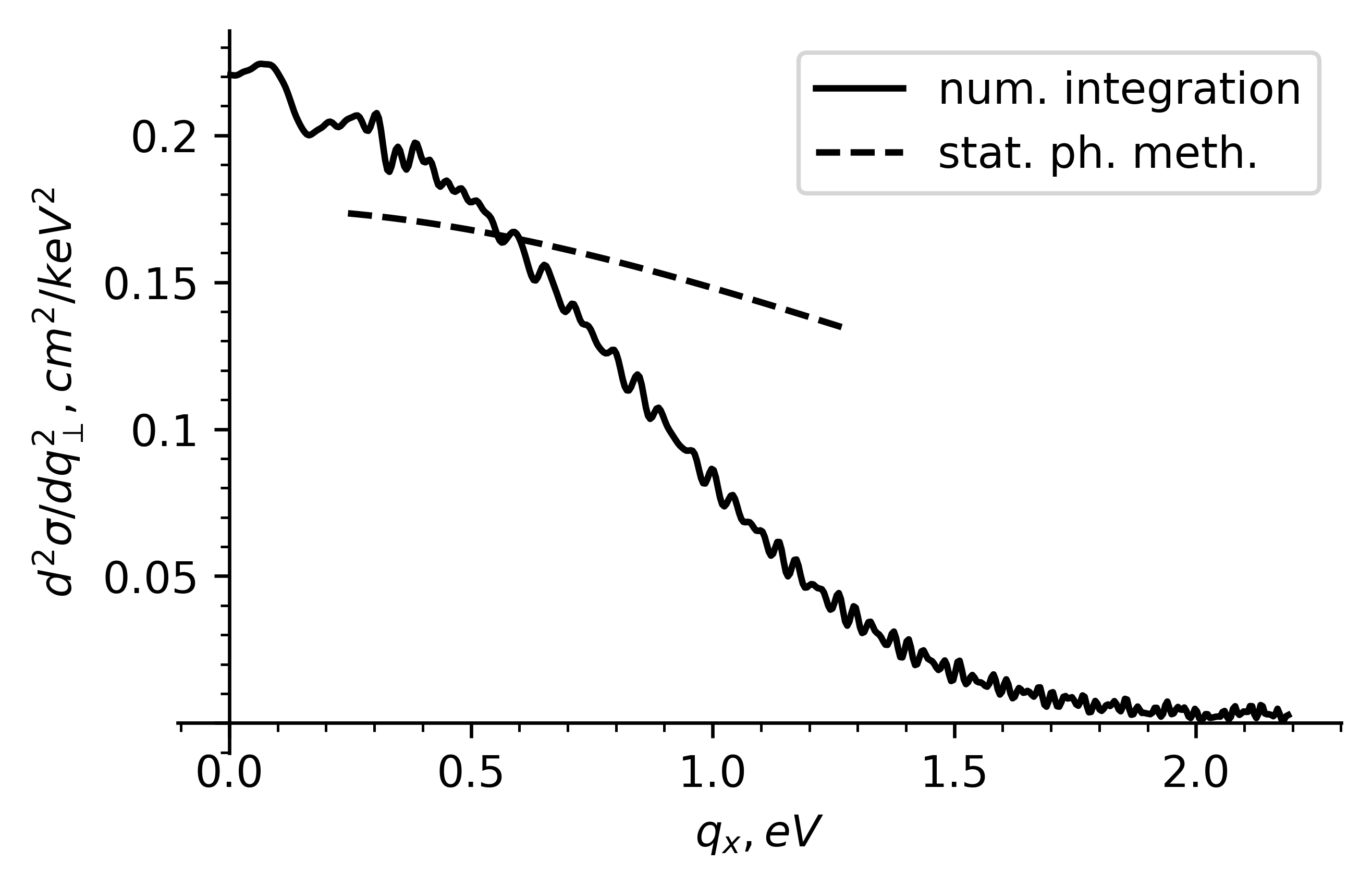}
		  \caption{$q_y=3.2 \ eV$}
		  \label{fig:subfig3_4}
	       \end{subfigure}
	\caption{The differential cross section for scattering on the flat beam of $N=10 \ 000$ particles and various $q_y$ values: solid line corresponds to numerical integration in \eqref{eq27}, dashed line -- to stationary phase method \eqref{eq31}}
	\label{fig:3}
\end{figure}
\par}
{The first derivatives for $x$ and $y$ are determined respectively
\begin{align}\label{eq29}
\partial_x \bar{\chi}_0^{(1)} =- \alpha \: \sqrt{\frac{2\pi}{\langle u^2_x \rangle}} \: Im \left\{ w \left[ \frac{x+i|y|}{\sqrt{2 \langle u^2_x \rangle}} \right] \right\}, 	
\end{align}
\begin{align}\label{eq30}
\partial_y \bar{\chi}_0^{(1)} = -\alpha \sqrt{\frac{{\pi}}{{2 \langle u^2_x \rangle}}} sgn(y) Re \left\{w \left[ \frac{x+i|y|}{\sqrt{2 \langle u^2_x \rangle}} \right] \right\},	
\end{align}
where $w[z]$ is Faddeeva function ($w[z]=e^{-z^2} Erfc \left[-iz \right]$, where $Erfc$ is complementary error function). \par}

{According to the stationary phase method, \eqref{eq27} can be approximately written as
\begin{align}\label{eq31}
\langle \frac{d^2\sigma}{dq_x dq_y} \rangle=\left. \frac{1}{N^2\left|\partial_x^2\bar{\chi}_0^{(1)} \partial_y^2\bar{\chi}_0^{(1)} - \left(\partial_{xy}\bar{\chi}_0^{(1)} \right)^2 \right|} \right|_{x=\tilde{x}, y=\tilde{y}},	
\end{align}
where the points $\tilde{x}, \tilde{y}$, which correspond to the phase minimum condition, are given by the system of equations
\begin{align}\label{eq32}
\begin{cases} q_x+N \partial_x\bar{\chi}_0^{(1)}=0, \\ 
q_y+N \partial_y\bar{\chi}_0^{(1)}=0.  \end{cases}	
\end{align}
The given system of equations corresponds to the quasi-classical approximation. According to \eqref{eq32}, the first derivatives of the $\bar{\chi}_0^{(1)}$-function determine the momentums transferred along the corresponding coordinate axes (we denote maximum absolute values of $q_x, q_y$ appropriate for this system as $q_{x \ max}, q_{y \ max}$). We will solve this system numerically. Let us illustrate the solution of this system graphically with the figure \ref{fig:0}: the solution is the point of intersection of the curves which correspond to the fixed  level lines of $\partial_x \bar{\chi}_0^{(1)}, \partial_y \bar{\chi }_0^{(1)}$. We can see that for some values of the momentum transfer the system of equations has no solution (for example, like in the figure \ref{fig:0} (b)). \par}

{Let us obtain results using the formula \eqref{eq31} for the region of momentum transfer, for which the system of equations \eqref{eq32} has a solution. We also calculate the differential scattering cross section by numerical integration of the formula \eqref{eq27}. Differential cross sections obtained with these methods are presented in figures \ref{fig:1}-\ref{fig:3}.
 \par}	
\section{Discussion}
{So, we obtained the differential cross section for a charged ultrarelativistic particle scattering on a flat beam of charged ultrarelativistic particles in the eikonal approximation of quantum electrodynamics. The eikonal approximation, compared to the frequently used Born approximation, has a wider application region, which justifies its usage. Also, to simplify the calculations, we used the continuous potential approximation. This approximation does not take into account incoherent scattering, but describes scattering in general terms. \par}
{As we see from the figures \ref{fig:1}-\ref{fig:3}, differential scattering cross sections obtained using numerical integration and using stationary phase method are of the same order of magnitude. The cutoff we made led to an error in numerical calculations for small momentum transfer. The stationary phase method has its own inaccuracy since it is an approximation and also gives a result only for the limited range of transferred momentums, unlike the numerical integration method which works even for classically forbidden region. Let us consider a "middle region" of transferred momentums where the momentum transfer is not small and at the same time is less enough than the maximum classically allowed momentum transfer. In this region, stationary phase and numerical differential cross sections agree quite well. This region of agreement enlarges with increasing number of particles in the target beam. As expected the accuracy of the stationary phase method increases when the phase of the wave function grows as $N$ increases. We also pay attention to the notion that the oscillations of the differential cross section obtained by direct numerical integration can be explained by the cutoff that we made. This issue requires additional consideration.  \par}
{We note that due to the long-range potential, the functions determining the scattering in this case have different properties compared to the functions associated with the short-range potential (which, for example, occurs in a crystal). The $\bar{\chi}_0^{(1)}$-function has a logarithmic divergence which is usual for the Coulomb potential, due to which we needed to implement the cutoff for physical reasons. The cutoff is justified by the fact that in real experimental facilities, the beam potential is screened at certain distances either by oppositely charged particles of the incident beam or by the walls of the facility. The  $\partial_x \bar{\chi}_0^{(1)}$ function, which according to the quasi-classical approach determines the momentum transferred along the $x$-axis, is proportional to $x^{-1}$ as $x$ approaches infinity, which means that it decreases much slowlier than a similar function for scattering in a crystal. Such behavior of the functions determines the significant influence of the long-range impact parameters on the scattering. It is unlike the scattering in short-range potentials, where the long-range impact parameters almost do not contribute to the scattering cross section. \par}
{The manifestation of the rainbow scattering effect for this problem also needs additional research, but we assume that for fixed transferred momentum $q_y$ and fixed impact parameters in the $y$-axis direction, the rainbow scattering is possible. This assumption is based on the form of the $\partial_x \bar{\chi}_0^{(1)}$-function (fig. \ref{FIG:5}), which determines the transferred momentum $q_x$ according to the quasi-classical approach.

\begin{figure}[!ht]
	\centering
			  \includegraphics[width=0.6\textwidth]{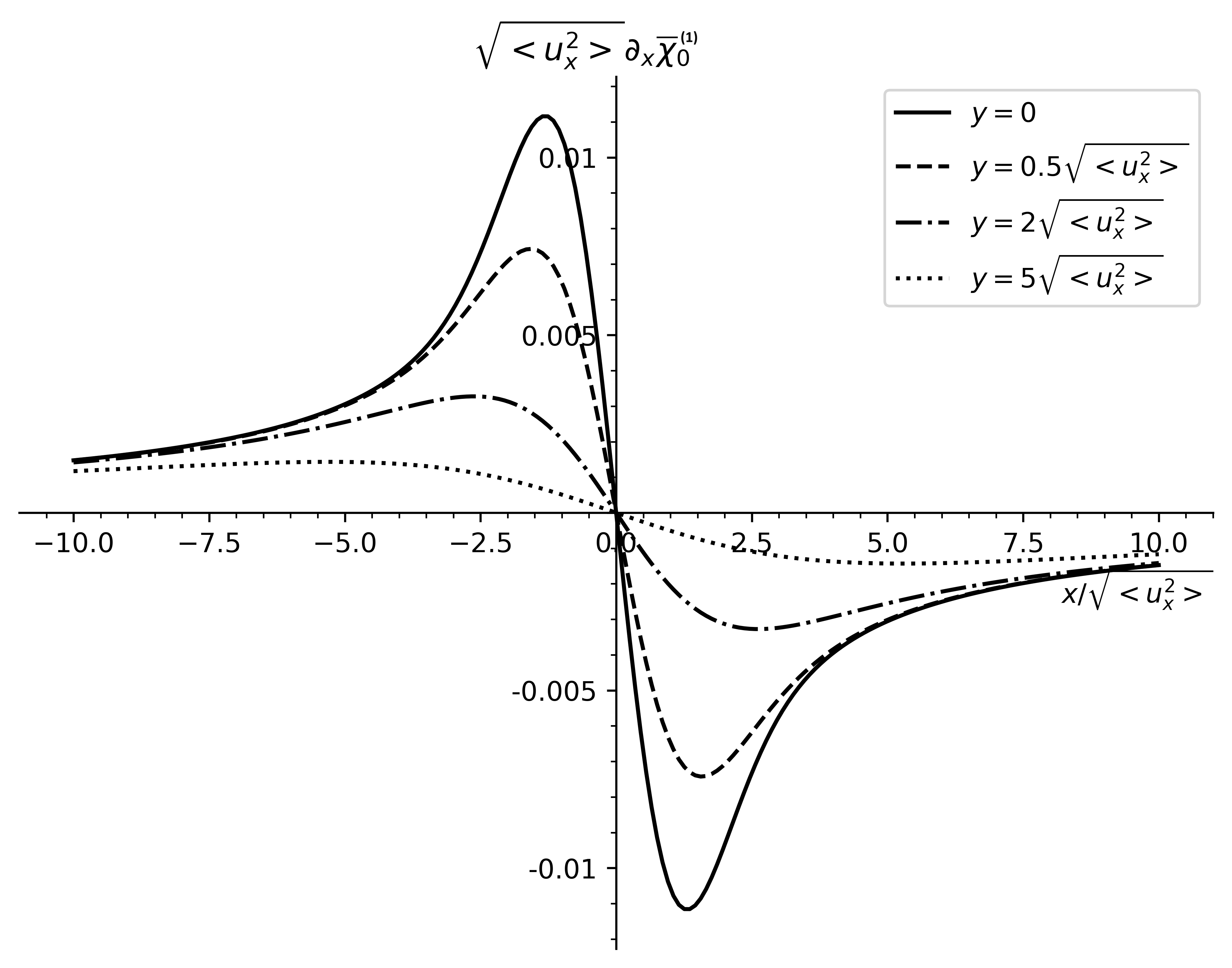}
	\caption{The $\partial_x \bar{\chi}_0^{(1)}$-function for the flat beam for various $y$ values}
	\label{FIG:5}
\end{figure}

 \par}

\section*{Acknowledgements}
{We thank God and His Blessed Mother for saving us and our Ukraine. The work was partially supported by the National Academy of \ Sciences of Ukraine (project 0123U103077), by Agence Universitaire de la Francophonie and by Wolfgang Pauli Institute. We also acknowledge the fruitful discussions with Kyryllin I.V., Trofymenko S.V., Bondarenko M.V., Omelchenko I.V.  
\par}

\end{document}